
\input phyzzx

\catcode`@=11 

\def\etal{{\it et al.}}

\def\kms{{\rm\, km/s}}
\def\kpc{{\rm\, kpc}}

\def\ten#1{\times 10^{#1}}
\def\msun {\, {\rm M_\odot}}

\def\frac{{f}}

\def\Amax{A_{\rm max}}
\def\umin{u_{\rm min}}
\def\tmax{t_{\rm max}}
\def\eff {{\cal E}}
\def\that{\hat{t}}

\def\pac{Paczy{\'n}ski}

\def\pri{^{\, \prime }}

\def\lsim{\mathrel{\mathpalette\@versim<}}
\def\gsim{\mathrel{\mathpalette\@versim>}}
\def\@versim#1#2{\lower0.2ex\vbox{\baselineskip\z@skip\lineskip\z@skip
  \lineskiplimit\z@\ialign{$\m@th#1\hfil##\hfil$\crcr#2\crcr\sim\crcr}}}
%
\def\spose#1{\hbox to 0pt{#1\hss}}
\def\simlt{\mathrel{\spose{\lower 3pt\hbox{$\mathchar"218$}}
     \raise 2.0pt\hbox{$\mathchar"13C$}}}
\def\simgt{\mathrel{\spose{\lower 3pt\hbox{$\mathchar"218$}}
     \raise 2.0pt\hbox{$\mathchar"13E$}}}

\PHYSREV
\singlespace
\pubtype{}
\pubnum{}
\hsize = 6.5truein  
\vsize = 9.0truein

\titlepage
\singlespace
\hsize = 6.5truein  
\vsize = 9.0truein

\title{\seventeenbf The MACHO Project: 45 Candidate Microlensing Events
                    from the First Year Galactic Bulge Data}

{\bf
\centerline {C.  Alcock$^{1,2}$, R.A.  Allsman$^3$, \
  T.S. Axelrod$^{1,4}$, D.P. Bennett$^{1,2}$, }
\centerline {K.H. Cook$^{1,2}$,
  K.C. Freeman$^4$, K. Griest$^{2,5}$, J. Guern $^{2,5}$, }
\centerline {M.J. Lehner$^{2,5}$, S.L. Marshall$^{2,6}$,H.-S. Park$^1$,
S. Perlmutter$^2$,
}
\centerline { B.A. Peterson$^4$, M.R. Pratt$^{2,6}$,
P.J. Quinn$^4$, A.W. Rodgers$^4$}
\centerline { C.W. Stubbs$^{2,6,7}$, W. Sutherland$^{8}$}
\centerline { (The MACHO Collaboration) }}
\vskip 0.25truein

\centerline{$1: $ Lawrence Livermore National Laboratory, Livermore, CA 94550}
\vskip 2pt
\centerline{$2: $ Center for Particle Astrophysics, }
\centerline{University of California, Berkeley, CA 94720}
\vskip  2pt
\centerline{$3: $ Supercomputing Facility, }
\vskip -2pt
\centerline{Australian National University, Canberra, A.C.T. 0200, Australia}
\vskip  2pt
\centerline{$4: $ Mt.  Stromlo and Siding Spring Observatories,}
\vskip -2pt
\centerline{Australian National University, Weston, A.C.T. 2611, Australia}
\vskip 2pt
\centerline{$5: $ Department of Physics, University of California, \
San Diego, CA 92093 }
\vskip 2pt
\centerline{$6: $ Department of Physics, University of California, \
Santa Barbara, CA 93106 }
\vskip 2pt
\centerline{$7: $ Departments of Astronomy and Physics,}
\centerline{ University of Washington, Seattle, WA 98195}
\vskip 2pt
\centerline{$8: $  Department of Physics, University of Oxford,
Oxford OX1 3RH, U.K.}

\vskip 0.2truein


\vskip 0.2truein
\centerline{ Subject headings: Microlensing;  Galaxy - structure;  Brown
   dwarfs }
\endpage

\abstract
\singlespace

We report the detection of 45 candidate microlensing events in fields
toward the Galactic bulge.  These come from the analysis of 24 fields
containing 12.6 million stars observed for 190 days in 1993.
Many of these events are of extremely high signal to noise and are
remarkable examples of gravitational microlensing.  The distribution
of peak magnifications is shown to be consistent with the
microlensing interpretation of these events.
Using a sub-sample of 1.3 million ``Clump Giant" stars
whose distance and detection efficiency are well known,
we find 13 events and estimate the microlensing
optical depth toward the Galactic Bulge as
$\tau_{\rm bulge} = 3.9 {+ 1.8 \atop - 1.2} \times 10^{-6}$ averaged
over an area of $\sim 12$ square degrees centered at
Galactic coordinates $\ell = 2.55^\circ$ and
$b = -3.64^\circ$. This is similar to the
value reported by the OGLE collaboration, and is marginally higher than
current theoretical models for $\tau_{\rm bulge}$. The optical depth
is also seen to increase significantly for decreasing $\vert b\vert$. These
results demonstrate that obtaining large numbers of microlensing
events toward the Galactic bulge is feasible, and that the study of
such events will have important consequences for the structure of the
Galaxy and its dark halo.

\endpage

\FIG\fieldfig{The locations of the 24 bulge fields presented in this paper are
    shown in Galactic coordinates. The dark spots indicate the locations of the
    events, and the area of each spot is proportional to the $\that$ value for
    each event.}
\FIG\cutfig{A scatter plot of the microlensing fit $\chi^2$ (per d.o.f.)
    vs.~$\Delta \chi^2$ which is the difference between the
    $\chi^2_{\rm const}$ for a constant light curve fit and the microlensing
    fit $\chi^2_{\rm ml}$. The solid circles indicate stars classified as
    clump giants while the $\times$'s indicate stars outside the clump
    giant region of the color magnitude diagram.
    The squares indicate events which are removed by cuts on parameters
    not displayed in this figure.}
\FIG\lcfig{Light curves for the MACHO blue and red pass bands are shown for
    the 45 events discussed in the text. The units are linear flux units
    normalized to the best fit unamplified flux given in Table 1.}
\FIG\cmdfig{The color magnitude diagram for each of the 45 event source stars
    along with a random selection of 7\% of the stars
    in $5^{\pri} \times 5^{\pri}$ regions containing each of these stars. The
    unamplified locations of the lensed stars are indicated, and the `clump
    giant' region defined in the text is shown as the region enclosed in the
    dashed contour.}
\FIG\efffig{The solid curve gives the sampling efficiencies for all stars
    while the dashed curve indicates the sampling efficiencies for the
    clump giant subsample. The clump giant sampling efficiency is a good
    approximation for the full efficiencies (including blending) for the clump
    giant sample.}
\FIG\uminKSfig{The cumulative distributions of the impact parameter, $\umin$,
    are compared to the predicted distributions for the full sample of events
    (a) and the clump giant subsample (b).}
\FIG\thatfig{A histogram of the $\that$ distribution of the 41 events
    used for the optical depth calculations.}
\FIG\taufig{A histogram of the $\that$ distribution with each event weighted
    by its value of $\that$. This is roughly the same as weighting each
    event by its contribution to $\tau$.}
\FIG\thatcncfig{The cumulative $\that$ distributions for the clump giant and
    non-clump-giant subsamples are plotted. The Kolmogorov-Smirnov (KS)
    test indicates that they are unlikely to have been drawn from the
    same distribution at the 0.7\% confidence level.}
\FIG\Evansfig{The cumulative distribution for the clump giant subsample is
    compared to the theoretical prediction of the evens Evans model. The
    KS test can be used to formally exclude the Evans model at 0.09\%
    confidence.}
\FIG\thatmodfig{Timescale distributions for 43/41  observed events (thick
    histogram) compared to predicted distributions, normalized to the
    observed number of events, for a fixed bar + disk density model (see text)
    with various mass functions.  The two dashed lines show delta-functions at
    $0.1 \msun$ (left) and $1 \msun$ (right); the solid line is the Scalo
    PDMF; the long-dashed line is the power-law with $\alpha = -2.3$,
    $m_{lo} = 0.1 \msun$ of Han \& Gould, and the dotted line is the `brown
    dwarf rich' mass function with $\alpha = -2$, $m_{lo} = 0.01 \msun$. }

\hsize = 6.5truein  
\vsize = 9.0truein


\singlespace

\chapter{Introduction }

The main goal of the MACHO project gravitational microlensing survey is to
search for massive compact halo objects (Machos)
  in the Milky Way halo in the mass range
$10^{-7}$--$100 \msun$, and thus to determine the contribution of
Machos to the mass of the Milky Way halo.
This is most readily achieved by searching for microlensing towards
the Magellanic Clouds (\pac\ 1986), where the microlensing rate
from a Macho-dominated halo is expected to be much larger than that from
known stars (e.g. Gould, Miralda-Escude \& Bahcall 1994,
Alcock \etal\ 1995d).

However, it was realised at an early stage in the project
 that the Galactic bulge is
also a promising target for microlensing surveys, for several reasons:
firstly, there is a ``known'' microlensing rate from dim stars in the
galactic disk and bulge (Griest \etal\ 1991; \pac\ 1991).
 Since there was considerable scepticism
that microlensing events could be detected, due to their rarity
and the need to reject intrinsic variable stars,
this provides a useful check on the experiments.
Also, since the duration of a microlensing event
is related to the mass of the lensing object,
microlensing towards the bulge can provide estimates of
the very low mass end of the stellar mass function, which is
difficult to measure directly, and can also test for the controversial
``disk dark matter'' (e.g. Bahcall 1986; Kuijken \& Gilmore 1989),
which (if it exists) is very likely to be in baryonic form.
 For a dedicated search
like the MACHO project, the Galactic bulge is also convenient from an
observational standpoint as it is visible when the Magellanic Clouds are
too low in the sky to observe.

When the first microlensing data from the Galactic bulge was analyzed, it
was discovered that the microlensing optical depth toward the bulge was
larger than expected (Alcock \etal\ 1995a, Udalski \etal\ 1994a) suggesting
that the standard models of the Galaxy needed to be revised. In fact, one
possible explanation of the large optical depth toward the bulge is that
the mass of the Galactic disk and bulge in ordinary stars is large enough
to account for almost all of the mass interior to the Sun. If true,
this would imply that the Galactic halo must have a large core radius or
perhaps a rather small total mass. Thus, contrary to expectations, microlensing
toward the Galactic bulge seems likely to reveal information about the
properties of the Galaxy that are important for determining the properties
of the Galaxy's dark halo.

In this paper, we present the results from the analysis of 1993 data from
our 24 well sampled Galactic bulge fields. We have searched this
data set for microlensing events and found 45 candidate microlensing events.
In \S~2 we review the basic physics and status of microlensing experiments,
in \S~3 we discuss our observations and photometric reductions, and
in \S~4 we discuss our automated search for microlensing
events, and the resulting candidates.
In \S~5 we provide an outline of the Monte-Carlo
simulations used to estimate our detection efficiencies.
In \S~6 we compare our observed distribution of peak magnifications
with the theoretical expectations and show that it is consistent with
the microlensing interpretation.
In \S~7 we make several estimates of
the optical depth of microlensing and compare our results
with predicted event rates from simple models of the galactic
disk and bulge. In particular, we
consider the subset of stars classified as clump giants and
obtain estimates of the optical depth which are relatively
free of systematic uncertainties.
In \S~8 we discuss the microlensing
event timescales, and
in \S~9 we discuss the implications of our observed microlensing
of a couple of bright main sequence stars.
In \S~10 and \S~11 we discuss the implications of
of these results for the composition of the Dark Halo and
summarize our conclusions.

\chapter{Microlensing}

The principle behind microlensing surveys is simple:
if a compact object (e.g. a Macho or faint star) passes very close to the
line of sight to a background star, the gravitational field of the
object deflects the starlight and produces multiple images of the source.
In the case of perfect alignment, the source star will appear as
an `Einstein ring', with a radius, $r_E$ in the lens plane defined by

$$ \eqalign{
r_E & = \sqrt{ 4 G m L x (1-x) \over c^2 }  \cr
     & = 2.85 \, {\rm AU} \sqrt{ \left({m \over \msun}\right) \left({L x (1-x)
\over 1 \kpc}\right) , } }
\eqn\rEeqn$$
where $m$ is the lens mass, $L$ is the observer-star distance, and
 $x$ is the ratio of the observer-lens and observer-star distances.
In a realistic case of imperfect alignment, the star will appear
as two small arcs. For the scales of interest here,
the image separation is $\simlt 0.001$ arcsec, and is far too
small to be resolved; however, the multiple imaging results in
an apparent amplification of the source (e.g. Refsdal 1964) by a factor
$$ A = {u^2 + 2 \over u \sqrt{u^2 + 4} }\ ,
\eqn\Aeqn$$
where $u = b/r_E$ and $b$ is the distance of the lens from the
undeflected observer-star line.
Since objects in the Galaxy are in relative motion, this
amplification will be transient, with a duration
$\that \equiv 2\, r_E / v_{\perp} $,
where $v_{\perp}$ is the transverse velocity of the lens relative
to the (moving) line of sight. For lens masses between a Jupiter mass
and a Solar mass, $\that$ is between a few days and a few months for most
Galactic populations.

The suggestion by \pac\ (1986) that microlensing could be used to search
for brown dwarfs or Jupiters that might comprise the Galactic halo
\footnote*{ A similar calculation was carried out by Petrou (1981), but was
not published.}
served to generate great interest in this technique and eventually
led to the microlensing survey projects which began a few years later
(Alcock \etal\ 1993, Aubourg \etal\ 1993, Udalski \etal\ 1993).

The great challenge of microlensing searches is that the microlensing
probability is very small when the lens is
in our Galaxy.
The ``optical depth'' $\tau$ to microlensing is defined as the probability
 that any given star is microlensed with impact parameter $u < 1$ (i.e.
 $A > 1.34$)
at any given time.
Since $r_E \propto \sqrt{m}$, while (for a given mass density)
the number density of lenses
$\overline{n} \propto m^{-1}$, $\tau$ is independent
of the mass function of the lenses,  and is given by
$$ \tau = {4 \pi G \over c^2} \int_0^L \rho_{lens}(l) {l (L-l) \over L} dl ,
\eqn\taudef $$
where $l$ is the distance to the lens, $L$ is the distance to the source
stars, and $\rho_{lens}(l)$ is the mass density of the lensing objects.
One complication for the optical depth towards the Galactic Bulge is
that the source stars are spread over a fairly large range of distances.
In this case, one must also average the optical depth
over the distribution of source star distances $L$.

To get an order of magnitude estimate of $\tau$, let us substitute
$M_{Galaxy}/L^3$ for $\rho_{lens}$ and identify $L$ with a `typical' Galactic
distance. Dropping all numerical factors, eq.~\taudef\ becomes
$\tau \sim GM_{Galaxy}/Lc^2\sim v_c^2/c^2 \approx 10^{-6}$ where $v_c$ is the
Galaxy's rotation speed. (We have used the virial theorem, $GM_{Galaxy}/L \sim
v_c^2$, to obtain this expression.) Thus, the optical depth for lensing
by objects in our Galaxy is of order $10^{-6}$ as shown by \pac\ (1986).

Although this optical depth is much lower than the fraction of
intrinsic variable stars ($\sim 0.3\%$), microlensing
events have many strong signatures which differ from all currently
known types of variable star. For microlensing events
involving a single point source, single lens, and uniform motions,
 the events are symmetrical and achromatic,
with a shape given by
$$ \eqalign{ A(t) & = A(u(t)) \cr
u(t) & = \left[ \umin^2 + \left({2 (t - \tmax ) \over \that }\right)^2
  \right]^{0.5} , }
\eqn\ueqn $$
where $A(u)$ is given by eq.~\Aeqn,  and $\Amax = A(\umin)$.
Since the optical depth is so low, only one event should occur in any
given star. If many events are found, additional statistical tests
can be applied:  the events should have a known distribution of
peak amplifications, they should be spread appropriately across
the color magnitude diagram, and the event timescales and peak amplifications
should be statistically independent.

To our knowledge, four groups have reported detections
of candidate microlensing events.
Our MACHO collaboration has reported four
candidate events towards the LMC (Alcock \etal\ 1993, 1994, 1995d),
while the EROS collaboration has reported two events towards the LMC
(Aubourg \etal\ 1993). Towards the Galactic bulge the event totals are
much larger: the OGLE collaboration reports
a total of 12 events (Udalski \etal\
1993,1994a),
and the DUO Collaboration has found about 10 events (Alard \etal\
1995, private communication). We have previously reported 4 of the 45 events
presented here (Alcock \etal\ 1995a), and our real-time ``Alert" system has
detected over 40 events toward
the Galactic Bulge during 1995\rlap.
\footnote\dag{Current information on the MACHO
Collaboration's Alert events is maintained at the WWW site:\hfil\break
{\tt http://darkstar.astro.washington.edu}, while the OGLE collaboration
maintains similar information at
{\tt http://sirius.astrouw.edu.pl/\~{ }ftp/ogle/ews.html.}}

\chapter{Observations and Photometric Reductions}

The MACHO project has full-time use of the $1.27$-meter telescope at
Mount Stromlo Observatory, Australia from mid-1992 through 2000.
The telescope was recommissioned especially for
this project, and a computer-controlled pointing and drive system was
installed.
A system of corrective optics has been installed
near the prime focus, giving a focal reduction to $f/3.9$
with a $1^o$ diameter field of view.
A dichroic beamsplitter and filters provide simultaneous
images in two passbands, a `red' band (approx. 6300--7600 \AA)
and a `blue' band (approx. 4500--6300 \AA).
Two very large CCD cameras are employed at the two foci;
each contains a $2 \times 2$ mosaic of $2048 \times 2048$ pixel
Loral CCD imagers. One half of one of the red focal plane CCDs does not
function.  The pixel size is $15\, \mu m$ which corresponds
to $0.63''$ on the sky, giving a sky coverage
of $0.72 \times 0.72$ degrees.
Each chip has two read-out amplifiers,
and the images are read out through a 16-channel system and
written into dual-ported memory in the data acquisition computer.
The readout time is 70 seconds per image, and the noise is
$\sim 10$ electrons rms, with a gain of $\sim 1.9 \, e^-$/ADU;
the images are written to disk and then saved on Exabyte tape.
Details of the camera system are given by Stubbs \etal\ (1993)
and Marshall \etal\ (1994).

Observations are obtained during all clear nights and partial nights,
except for occasional gaps for telescope maintenance.
The default exposure times are 300 seconds for LMC images, 600
sec for the SMC
and 150 seconds for the bulge, so over 60 exposures are
taken per clear night.
As of 1995 August, over 35000 exposures have been taken with the system,
of which about 60\% are of the LMC,  10\% of the SMC and 30\% of the bulge.
The images are taken at standard sky positions, of which we have defined
82 in the LMC, 21 in the SMC and 75 in the bulge
\footnote\ddag{
Coordinates of the field centers are available on the WWW, \hfill
 \break URL: {\tt http://wwwmacho.anu.edu.au}}

Because the primary goal of the MACHO project is to measure the
density of Machos residing in the Galactic halo, observations of the
LMC and SMC have been given priority, so the bulge is observed only when
the LMC is at an elevation $\lsim 25^\circ$.
In the first season of bulge observations in 1993,
we concentrated on observing
a subset of our bulge fields which is relatively close to the Galactic
plane. During much of the observing season, many of the bulge fields
were observed twice per night to improve the sensitivity to short
time-scale events.

In this paper, we consider {\bf only} the 1993 data from 24
well-sampled bulge fields containing dual color lightcurves for
12.6 million stars. The 24 fields used for this analysis are
Macho field numbers
101, 102, 103, 104, 105, 108, 109, 110, 111, 113, 114, 115, 118, 119, 120,
121, 124, 125, 128, 159, 161, 162, and 167. The positions of
these fields are indicated in Figure \fieldfig.
The observations analysed here comprise 2313 images,
covering a time span of 189 days from 1993 February 27 to 1993 September 03.
The mean number of exposures per field is $2313/24 = 96$
with a range from 54 to 165.
This sampling varies quite substantially among our fields, since we
usually observed the fields in a fixed order each night so that our
``highest priority'' fields were always observed even
on partially clear nights, and were frequently observed twice per night.

\section{Photometric Reductions}
Photometric measurements from these images are made with a
special-purpose
code known as SoDoPHOT (Bennett \etal\ 1996), derived from DoPHOT
(Schechter \etal\ 1993).
First, one image of each field with good seeing and dark sky
is chosen as a `template image'.
This is processed in a manner similar to
a standard DoPHOT reduction except that after one color of the
image has been
reduced, the coordinates of the stars found in the first color are used
as starting points for the positions of stars in the second color;
this improves the star matching between colors.
(The final positions of the matched stars are forced to
be the same in both colors,
after allowing for differential refraction.)
This procedure provides a `template' catalog of stellar
positions and magnitudes for each field.

All other images are processed in `routine' mode, which proceeds
as follows. First the image is divided into 120 `chunks' of
$\sim 512 \times 512 $ pixels, and for each chunk
$\sim 30$ bright stars are located and matched with the template.
These stars are used to determine an analytic fit to the point spread
function, a coordinate transformation,  and a photometric zero point
relative to the template.
Then, all the template stars are subtracted from the image using
the model PSF and coordinate transformation, and noise is added to the
variance estimate for each pixel to allow for errors in the subtraction.
Next, photometric fitting is carried out for each star in descending
order of brightness, by adding the analytic model of the star
back to the subtracted frame and fitting a 2-parameter model of the stellar
profile and sky background, with pixels weighted by inverse variance,
while the model PSF and computed position of the star are kept fixed.
When a star is found to vary significantly from its template magnitude,
it and its neighbors undergo a second iteration of fitting.
For each star, the estimated magnitude and error are determined,
along with 6 other parameters measuring the object `type',
the $\chi_{\rm PSF}^2$ of the PSF fit, the crowding, the weighted fractions
of flux removed due to bad pixels ($f_{\rm mis}$) and cosmic rays
($f_{\rm CR}$), and the fitted sky value. The crowding parameter $f_{CRD}$
is defined to be the ratio of the flux contributed by other stars
over the flux contributed by the individual star
to that star's central pixel, in seeing 30\% worse than the actual seeing.
The photometric error estimate is the formal PSF fit error (as in
DoPHOT) with a 1.4\% systematic error added in quadrature.
These routine reductions are completed by SoDoPHOT at a rate of
approximately 1 million photometric measurements per hour on a Sparc-10.
The set of photometric data points
for each field are re-arranged into a time-series for each star,
combined with other relevant information including the seeing
and sky brightness, and then
passed to an automated
analysis to search for variable stars and microlensing candidates.

\chapter{Event Detection}

The analysis of microlensing survey data presents some unusual challenges.
The microlensing signal that we seek to detect affects only a few stars
per million while a few stars per thousand are intrinsically variable.
The vast majority of these variable stars are of known types and do not remain
at a constant brightness for long periods of time. These common types
of variable stars are not
easily confused with microlensing. However, there may exist unusual
variable star types which do resemble microlensing events, and these
would be an important background for the microlensing search.
Because microlensing surveys
are, by far, the largest scale searches for stellar variability to date,
we should not expect to learn about possible microlensing-like variable
stars except through our own data. This complicates microlensing event
detection.

In a laboratory experiments, one typically deals with the background by
modeling it and then either subtracting or fitting the background
model to the data. In our case, however, the variable star background
is not know well enough to be modeled. Instead, the background
must be discovered and characterized using the same data set that
we use to search for microlensing events. The situation is not as bad
as one might think because the high amplification subset of the microlensing
events have lightcurves (\eg events 101-D and 108-D in Figure \lcfig)
that are qualitatively very different from any variable star ever observed.
Nevertheless, the event detection procedure is somewhat subjective, and some
care must be taken to avoid biasing the results during the event
detection analysis. We have attempted to do this using a set of cuts
that are either fairly simple or have been determined {\it a priori}
based upon analysis of an independent data set. For this analysis, all cuts
which did not involve the $\chi^2$ of the microlensing fit have been
determined {\it a priori} while the cuts involving the microlensing
fit $\chi^2$ have been adjusted to fit the characteristics of this data set
and the variable star background seen toward the bulge. The number of
events close to the cut boundaries is small (especially for the clump
giant sub-sample), so the conclusions reached below do not depend on the
cuts we have chosen.

The first stage of the microlensing search is to define the set of
`acceptable' data points, using the
PSF chi square, crowding, missing pixel, and cosmic ray flags
described above.
We have investigated the relationship between these quality flags
and apparent `bad' measurements as follows: we defined a set
of `non-variable' stars using a robust $\chi^2$ measure,
which is designed to reject periodic variables while including
stars with occasional discrepant data points. For these stars,
we then examine the percentiles of the
distribution of $\Delta m/ \sigma$ for many distinct bins of each flag.
As expected, data points with large values of the various flags
generally show a significant non-gaussian tail of outliers;
thus, we set the following cuts on
the various flags so as to reject most such outliers:
$f_{\rm CR} < 0.001$, $f_{\rm mis} < 0.004$, $f_{CRD} < 3.0$, and
$\chi^2_{\rm PSF}({\rm d.o.f.}) < 4.0$. For bright stars, the deficiencies
of the analytic PSF model become apparent, and this means that we must
turn off the cut on $\chi^2_{\rm PSF}({\rm d.o.f.})$ for stars with more
than 63,000 detected photoelectrons.
Data points failing any of these cuts are marked as `suspect';
they are retained in the database, but are not used in the
microlensing or variability searches.

We exclude the reddest $0.2\%$ of stars with $V-R > 1.6$
from the microlensing search as these are often long-period variables
which nearly always trigger the fitting routine,
and would dominate the overall number of triggers. Stars which are very
close to a chip boundary or which have less than
7 simultaneous red-blue measurements are also removed from the microlensing
search.

The microlensing search through the light curve database
proceeds in three stages: first, the time-series are convolved with a set
of microlensing lightcurve filters of durations
7, 15 and 30 days in order to search for peaks of any kind.
Any lightcurve with a significant peak in any filter is
tagged as a `level--1' trigger; about $1\%$ of the stars pass this
trigger. For these level--1 lightcurves
a 5-parameter fit to a microlensing event is made, where
the parameters are the un-amplified red and blue fluxes $f_{R0}$,$f_{B0}$,
the peak amplification $\Amax$, the time of peak amplification
$\tmax$ and the event timescale $\that$.
Thus, the fitted flux of the star in each color is given by
$$\eqalign{
f_B(t) & = f_{B0} A(u(t)),  \cr
f_R(t) & = f_{R0} A(u(t)),   \cr }
\eqn\fluxfac $$
where $A(u)$ is given by eq.~\Aeqn, and $\Amax = A(\umin)$.
Following the fit, a set of statistics
describing the significance level, goodness of fit, achromaticity,
crowding, temporal coverage of the event, etc. are calculated.
Events above a modest significance level are tagged as `level-1.5' events and
are output as ASCII files, along with their associated statistics;
these level-1.5 candidates are then
subjected to more rigorous selection criteria, which may be easily
modified, to search for final `level-2' microlensing candidates.

Out of the 12.6 million stars in
this data set, 37,485 stars passed the level-1.5 criteria.
The most important of the `level-2' cuts are shown in Fig.~\cutfig.
The x-axis of Fig.~\cutfig\ is
$\Delta\chi^2 \equiv \chi^2_{\rm const}-\chi^2_{\rm ml}$, the difference
between the $\chi^2$ values for the constant-flux fit and the
microlensing fit,
while the y-axis of Fig.~\cutfig\ is
$\chi^2_{\rm ml}({\rm d.o.f.})$ which is the $\chi^2$ per degree of
freedom for the microlensing fit.
The large number of events at low $\Delta\chi^2$ generally contain
small bumps attributable to low-level systematic errors.
More than 90\% of these level 1.5 candidates
are stars which are just barely resolved from their closest neighbor in
our best seeing images. They
have spurious lightcurve bumps between days 175 and 190 which corresponds
to a period when the telescope was out of alignment slightly because of
a mirror support problem which was later repaired. We believe that the
cause of the spurious photometry was the failure of the elliptic PSF
used by SoDoPHOT to fit an asymmetric PSF caused by the telescope
misalignment.
This failure would allow the photometry of stars with very close neighbors
to be contaminated with unsubtracted flux from their neighbors.

The two cuts shown in Fig.~\cutfig\ are
$\Delta\chi^2/\chi^2_{\rm ml}({\rm d.o.f.}) > 400$ and
$\Delta\chi^2/[\chi^2_{\rm ml}({\rm d.o.f.})]^2 > 200$, and they are
sufficient to cut the list of 37,485 `level 1.5' microlensing candidates
down to 52. Additional cuts
on crowding parameter ($\VEV{f_{CRD}}<1.67$), time coverage, and
the $\chi^2$ in the peak region ($\Delta\chi^2/\chi^2_{\rm peak} < 200 $)
remove the 8 events indicated by squares
to yield 42 candidate events which are summarized in Table 1.
(Three additional microlensing candidates which do not pass our cuts are
also included in Table 1.) Lightcurves
of these events are shown in Fig.~\lcfig\ along with the best fit
theoretical microlensing lightcurves.
Some noteworthy events are indicated with superscripts on the event ID
number in Table 1. A superscript ${}^{\rm c}$ indicates that the source
star is classified as a clump giant, a class that will assume a special
role when we estimate the microlensing optical depth.

The fits presented in Figure \lcfig\ do not account for the possibility that
the lensed star might be blended with one or more other stars in our images.
In this case only a fraction of the flux of an
object that our photometry code has
identified as a star will actually be lensed. In principle, one could account
for this by allowing for an unlensed source to be superimposed on the lensed
source in the microlensing fit. We have chosen not to do this for a number
of reasons. The main difficulty is that there is a near degeneracy in
parameter space for blended point mass microlensing light curves. By decreasing
the fraction of the source that is lensed as the peak amplitude and
duration are increased, one can construct a family of light curves with
very similar shapes. Also, the amount of blending typically depends on seeing
which varies significantly between observations so that stars that are
identified to be separate in good seeing images are blended in poor seeing.
In this situation, the photometry code preferentially selects the brighter
star of the blend to receive the additional flux when the seeing is poor
and the change in brightness is small. This can systematically change the
shapes of microlensing light curves and prevent a meaningful determination
of the blend fraction for a microlensing event.

In practice, we find that allowing for blending in our microlensing fits
yields physically unlikely solutions for a large fraction of our lensing
events. This suggests that artifacts like the one mentioned above are
influencing the fits. At present, we are developing a new photometry routine
(in collaboration with P. Stetson) that will reduce many images
of previously detected events simultaneously. This routine
should avoid the sort of artefacts mentioned above, so we will re-examine the
issue of blending when this improved photometry is available.

In the present paper, however, the reader should note that the 1-$\sigma$
fit errors reported in Table 1 tend to underestimate the true uncertainties
in these quantities. This is mostly due to blending, but correlated
and non-Gaussian measurement errors probably also contribute to this.

\section{Microlensing Events}

Table 1 lists 45 microlensing events which include the 42 events which
passed all of our cuts and 3 events which have been selected by eye as
likely microlensing events from a larger sample of events passing less
restrictive cuts. One of these three events is the binary lens
event first seen by the OGLE group (Udalski \etal\ 1994b).
In addition, we have determined that 2 of the 42 events which have passed
our cuts are probably not microlensing events, and these are not used
in our subsequent analysis. We do, however, include the binary lens
event in most of our analyses in order to partially compensate for the
fact that the microlensing fits used for event detection do not allow
for binary lenses.

Clearly, our candidate microlensing events span a wide range
in quality; some are of remarkably high signal-to-noise (e.g. 101-B,
108-D, 118-B) while others are relatively unimpressive (e.g. 110-C, 114-C).
This is just as expected due to the range of amplifications and
stellar magnitudes involved; we provide a detailed discussion of these
distributions in \S4.2 and \S6.

A number of the events shown in Figure~\lcfig\ warrant further discussion.
Three events have best fit amplifications substantially higher than the
highest measured point. For events 101-D and 124-B where the highest
measurement has $A\approx 15$, it is probably true that the best
peak amplification was significantly higher, but it is also possible that
the light curve of these stars may deviate from the form given by
eqs.~\Aeqn\ and \ueqn\ if $u(t)$ should become as small as the projected
radius of the source star. This would cause the light curve to steepen
near the peak and then flatten off more abruptly at the peak. Undersampled
events with this type of light curve will often generate best fit peak
amplifications which are substantially in excess of the highest measured
amplification. If events like 101-D, 124-B, and 108-D had been discovered
prior to peak amplification by our alert system (which only became operational
in August, 1994), then we would have the opportunity to obtain photometry with
much better time resolution. Photometry of high amplification events like
these with better time resolution would enable us to measure or set
an interesting limit on the projected Einstein ring radius by comparison with
the radius of the source star.
[This effect has now been detected in MACHO Alert 95-30, and will
be discussed in a later paper].

Two of the events (104-C and 119-A) shown in Figure~\lcfig\
exhibit exotic deviations of a different type.
Event 104-C shows a ``parallax" effect in which the motion of the earth has
caused $u(t)$ to deviate from the uniform motion assumed in
eq.~\ueqn. This results in a slight asymmetry of the lightcurve, causing
deviations from the best fit symmetric light curve shown in Figure \lcfig.
By fitting this light
curve with a model which takes the motion of the earth into account, we are
able to obtain a very good fit and thereby measure the lens velocity
projected to the solar position. Details of this analysis and its
implications are discussed in Alcock \etal (1995e).
The other `exotic' event is 119-A, which is a spectacular
example of microlensing by a binary lens. This event was
first seen by OGLE as event OGLE\#7 (Udalski \etal\ 1994b);
 our data has considerably better time coverage which strongly
confirms their binary lens interpretation.
In our data, the second caustic crossing is resolved which is the
first time such an effect has been observed (Alcock \etal\ 1995f).
Note that while event the parallax event (104-C) passes all our cuts, the
binary lens event (119-A) does not and must be moved from level-1.5 to
level-2 by hand. It is necessary to do this because we have not incorporated
binary lens fitting into our analysis.
These events will be discussed in more detail in subsequent papers.

The superscript ${}^{\rm f}$ in Table 1 is used to indicate the two events
which seem to be microlensing events upon inspection
but which do not pass our final level-2
cuts. One of these is event 104-B which fails the cut on the average value of
the crowding parameter. This cut is designed to help remove some of the
spurious level-1.5 microlens triggers which are caused by the effects of
asymmetric point spread functions
on very crowded stars. Unfortunately, it also has the
effect of removing one or two likely microlensing candidates such as 104-B.
The other probable event that fails the final cuts is 111-B. This
fails because the time of the fit peak occurs later than day 238.5,
which is the latest time for which we have added simulated events in the Monte
Carlo calculations used to estimate our detection efficiency, so we must
exclude it from our subsequent analysis which relies upon the efficiency
estimates.

Finally, there are two events (denoted in Table 1 with a ${}^{\rm v}$)
which pass the cuts, but which we believe are not actual microlensing
events. Event 113-C is located very close ($\sim 1"$) to a bright,
very red long period variable which appears to be in phase with the
event 113-C light curve. The behavior of the crowding parameters in the
113-C light curve suggests that the variation seen is probably due to
contaminating flux from the neighboring variable, and not
to star 113-C.
The other suspicious event
is 121-B. This light curve appears to resemble the light curves of a few
other stars which do not quite pass the cuts, in that it appears to have
a larger amplification in
our blue passband and is somewhat asymmetric. We feel that there
is a reasonably high probability that this event is an intrinsic variable
and not a microlensing event. These two events are not used in the
subsequent analysis.

In addition to these stars, there are a few
other microlensing candidates in Table 1 and Figure \lcfig\ which have
rather low signal to noise and/or poor time coverage. Because the vast
majority of events which pass our cuts appear to
have strong microlensing signatures, it appears that the microlensing
event `signal' is higher than the `background' of microlensing--like
variable stars. Thus, it seems likely that most of the `low quality' events
are also microlensing. We would like to emphasize, however, that our
principal conclusions do not depend on these `low quality' events.

Our field 119 covers about the same area of sky as the 9 OGLE fields
BW1--BW8 and BWC, so we should expect to have some events in common
with the OGLE collaboration. Unfortunately, the majority of their events
occurred in 1992, before we were taking data on the bulge,
but we confirm that all 5 of their 1992 lensing candidates
in our field 119 remained constant in our 1993 observing season.
There are 2 OGLE candidates in Baade's Window during 1993, and we
have rediscovered both of them: our event 119-A is OGLE-7 mentioned above,
and event 119-D is OGLE-1.
A third event seen by OGLE in 1993 (OGLE-8) falls outside our fields.
We have also discovered two new events in field 119, namely 119-B and
119-C: these do occur in the OGLE fields but were not found by OGLE.
A comparison with the light curves
of the other OGLE events suggests that these events occurred during
gaps in the OGLE observing schedule,
so we would not expect OGLE to have detected them.
We emphasize that this does not cast doubt
on any OGLE results, since such gaps are already accounted for in
their efficiency analysis (Udalski \etal\ 1994a).

In the following sections, we use several different samples of events.
The optical depth and timescale analyses use the sample of 41 events
which includes all events except for 104-B, 111-B, 113-C and 121-B. (The
events which fail the cuts or are likely to be due to photometry errors
and stellar variability are excluded.) For the distribution of peak
amplifications, events 104-B, 111-B, and 119-A (the binary lens event)
are excluded leaving a sample of 42 events. A proper binary lens fit
would be required before event 119-A could be included.

\section{Color Magnitude Diagram}

Figure \cmdfig\ shows a color magnitude diagram
showing the locations of the baseline colors and magnitudes for each of
these 45 stars as well as 7\% of the stars
in  $5^{\pri} \times 5^{\pri}$ regions containing each of these stars.
The color bands used for the plot are
\def\Vmacho{{\rm V_M}}
\def\Rmacho{{\rm R_M}}
$$ \eqalign{ \Vmacho &= 0.94 \, v + 0.06 \, r \cr
             \Rmacho &= 0.32 \, v + 0.68 \, r \cr }
\eqn\Cbandeqn $$
where $v$ and $r$ are the MACHO instrumental passbands, and
$\Vmacho$ and $\Rmacho$ are designed to approximate Johnson V and R.
No correction has been made for extinction, so the diagram is smeared out
by extinction along a direction that is parallel to the left face
of the quadrilateral drawn on the diagram. This quadrilateral is the
region which we select as the `clump giant' region of the color magnitude
diagram. This region is defined by:
$$ \eqalign{ 0.1076 \,\Vmacho - 1.008 &\leq \Vmacho - \Rmacho \leq
        1.6 \cr
    \rm \Rmacho &\leq 18.0 \cr
    \rm \Vmacho &\geq 16.0 \cr }
\eqn\clumpbound $$
The region that is really populated by genuine clump giants
(core helium burning horizontal branch stars) is the heavily populated
region of the diagram close to the diagonal left face of this region.
We have extended this region to include giants somewhat redder than the
clump giants themselves because these stars are also (mostly) located in
the bulge, and we would want to include any events found in this area of
the color magnitude diagram.

There are a few features in the distribution of events on the color magnitude
diagram that are of particular interest.
It is evident that the events are spread all over the
heavily populated regions of the color magnitude diagram, and this is
consistent with the hypothesis that a large fraction of the
events are truly microlensing;
other types of variable stars generally appear only in particular
regions of the color magnitude diagram. 
It is also true that the
microlensing events are not a representative sample of the color magnitude
diagram as there is a much larger fraction of microlensing events among the
bright red clump giant branch stars than among the much more numerous
fainter stars. There are two reasons why we expect this to be the case.
The first is that our microlensing detection efficiency is much larger
for the brighter stars because their signal--to--noise is higher
and because crowding problems are less severe for brighter stars.
Another reason is that the stars in different regions of the color magnitude
diagram do not have the same distribution of distances. For example, a
likely explanation of why there are so many more events on the
bright giant branch
than on the main sequence is that the giants are expected to reside mostly
in the Galactic bulge while the main sequence stars are expected to
be mostly foreground stars in the disk.
This would substantially lower the microlensing optical
depth for the main sequence stars.  We discuss this more in \S~9.

Another location in the color magnitude diagram which seems to be somewhat
over represented in the set of candidate lensing events are the very faint
stars at the bottom of the diagram. In particular, the three events 120-A,
159-B, and especially 110-C come from an extremely sparse region
at the bottom of the color magnitude diagram. There are several possible
explanation for this. One very important contribution is amplification
bias. Star 159-B was amplified by about 1.9 magnitudes in the template
observation, and star 110-C was amplified by 0.9 magnitudes. Star 110-C
probably would not have been above the detection threshold in the template
if it had not been amplified, but star 159-B probably would have since it is
in one of our less crowded bulge fields. Also, event 120-A and particularly
event 110-C are rather low signal-to-noise, it is possible that one or both
of these events are due to stellar variability rather than microlensing.
A final possibility to be considered is that some of these faint stars
may be significantly behind the bulge where the optical depth to microlensing
would be much higher.

\chapter{Microlensing Detection Efficiency}

Before we can draw any conclusions about the statistical properties of the
detected microlensing events, we need to assess our microlensing
detection efficiency. There are a number of effects that influence
our detection efficiency, and we separate them into two different
classes which we refer to as the sampling efficiency, $\eff_s$,
and the blend efficiency, $\eff_b$.
The detection probability for a single event is a function of
the event timescale $\that$, the stellar magnitude,
the peak amplification and the peak time,
but we can average over the latter 3 parameters using
the known distributions, giving our efficiencies as a function only of
event timescale $\eff_s(\that), \eff_b(\that)$.

The sampling efficiency, $\eff_s(\that)$, is defined as the fraction of
microlensing events in all monitored stars with $\Amax > 1.34$ which
we expect to detect, taking into account the actual spacing of the
observations, the variable seeing and sky brightness,
as well as our event selection criteria.
The sampling efficiency is calculated by adding a single simulated
microlensing event to a random $1\%$ of all stars,
with a uniform distribution in $\log \that$ for
 $ 0 \le \log_{10}(\that/{\rm days}) \le 2.5$, and uniform distributions in
$\umin$ and $\tmax$.
For each event and each data point, we add the excess flux given by
 $(A(t)-1) f_{\rm med}$ to the observed flux, where $A$ is the
theoretical amplification, and $f_{\rm med}$ is the median observed
flux of the star; this preserves the ``real'' scatter in the
data points.
We then search this `simulated event' lightcurve database using the
same procedure as in the actual lightcurve database, and
$\eff_s(\that)$ is defined as the fraction of events with
time scale $\that$ that are recovered.
Thus, the sampling efficiency takes
into account events that are missed because of imperfect time sampling,
periods of poor photometry due to bad seeing or a bright sky background,
or just because they fail some of the cuts on crowding, color, etc.
This method implicitly assumes
the database contains photometry of single (unblended) stars, and that the
photometric errors are well characterized by the error estimates that
are generated by the photometry code.
Figure \efffig\ shows the sampling efficiencies for all our stars,
and for only the stars falling within the clump giant region of the
color magnitude diagram (Figure \cmdfig).

In the crowded stellar fields that are surveyed
for microlensing events,
the blending of overlapping stellar images causes many of the fainter
stars not to be identified as individual stars. Furthermore, since the
positions of the blended stars do not, in general, exactly coincide, the
blending effects will depend on seeing. Blending makes it more difficult to
detect microlensing events because the blend of multiple stars is
amplified by a smaller amount than the lensed star itself. Of course,
one also has a chance to detect lensing of the other members of the blend,
which can partially counteract the decrease in sensitivity caused by the
blending, but it can be shown that a microlensing search is the most
sensitive in the case where all the detected stars are actually unblended
single stars.

We refer to the efficiency calculated
with the effects of blending taken into account as the blend
efficiency, $\eff_b$. This is properly determined by adding
``artificial" stars with the same luminosity function as the real stars to the
raw images. The brightnesses of these artificial stars can be modulated
according to randomly selected microlensing light curves, and then the
analysis can be run to see how many of these simulated events are recovered.
To date, the MACHO collaboration has calculated blend efficiencies only for
our first year LMC data analysis (Alcock \etal\ 1995c, 1995d).
There, we find that the effect of blending is to
reduce $\eff_b(\that)$ to be 20--40\% less than $\eff_s(\that)$
depending on $\that$. The bulge luminosity function is rather different from
that of the LMC, and our bulge fields have an average stellar density
some $30\%$ higher than in the LMC.
 Thus, the ratio of blending to sampling efficiencies in the bulge
may be somewhat different to that in the LMC, and we parameterize
this by the uncertain factor $f_{blend} = \eff_b / \eff_s \sim 0.75$.
In order to minimize this uncertainty , we will draw our main conclusions from
the subset of `clump giant' stars. These stars are much brighter than the
typical stars in these fields, so significant
blending is uncommon. Thus, the approximation $\eff_b \simeq \eff_s$
is a reasonable one for the `clump giants\rlap'.

One aspect of microlensing that we have not taken into account
is the fact that the standard microlensing light curve used
in our analysis and efficiency calculations does not describe all microlensing
events. The parallax and binary lens events (104-C and 119-A) are clear
illustrations of this fact. It is, of course, possible to include exotic
events in the simulated event light curves used to estimate our efficiencies,
but this involves assumptions about a number of new parameters that we
have little information about. In the present paper, we make the (crude)
assumption that our efficiency to detect exotic events is the same as
our efficiency to detect events which follow the standard microlensing
light curve. This causes us to overestimate our efficiency somewhat, but
since these exotic events are rare this is not likely to be a large effect.
This also justifies our inclusion of event 119-A in
our microlensing candidate list despite the fact that it did not pass
our cuts. Such a spectacular event would certainly pass any reasonable
set of cuts that takes binary events into account, so we clearly need to
include it to reduce the ``exotic event" error in our efficiency estimates.
We should also point out that this crude treatment of exotic lensing events
clearly implies that we can make no meaningful
statements about the fraction of binary
lenses toward the Galactic bulge based upon the current analysis.
We plan to address this point in a future publication after we have
completed a more comprehensive study of our detection efficiencies.

\chapter{Distribution of Peak Amplifications}

One important test of the microlensing hypothesis is that the distribution
of peak amplifications should follow the theoretical prediction
that actual events should be uniformly distributed in $\umin$.
In order to compare with the distribution of detected events, we must
include the fact that the detection efficiency depends on $\umin$.
Ideally, we should use the full blend efficiencies in the comparison of
the $\umin$ distribution, but as we have stated, these efficiencies are
not yet available (except for the subset of clump giant events where we have
argued that the full efficiency equals the sampling efficiency to a good
approximation).  In what follows we test the microlensing hypothesis using
sampling efficiencies both for our entire sample of microlensing events
and for the clump giant sub-sample.

The Kolmogorov-Smirnov (or K--S) test is a convenient statistical test
to compare the predicted and observed $\umin$ distributions. We will
compare the predicted to the observed $\umin$ distribution for the set of
42 events which pass our cuts. (The binary lens is {\it not} included
in this comparison.)
In order to correct the $\umin$ distribution for the
sampling efficiency, we want to average our calculated
sampling efficiency $\eff_s(\umin$,$\that$) over the underlying
$\that$ distribution.  An approximation to the actual
$\that$ distribution can be calculated by taking the observed $\that$
distribution and weighting each event by the inverse of its detection
efficiency.  Averaging over this distribution gives the $\umin$ distribution
we should expect in our observed data, given that the
actual $\umin$ distribution should be uniform.

The cumulative distribution functions are shown in Figure \uminKSfig(a),
where the solid line shows the efficiency corrected theoretical distribution.
The K--S statistic is 0.13 which gives $P_{KS} = 0.41$. We
also compare to the uncorrected theoretical distribution (the dashed line).
This curve is shown to indicate the effect of the sampling efficiencies on
the $\umin$ distribution. Since the difference between the full blend
efficiency ($\eff_b(\that)$) and the sampling efficiency
($\eff_s(\that)$) is much smaller than the difference
between $\eff_s(\that)$ and $\eff(\that) \equiv 1$, the comparison
to the theoretical $\umin$ distribution corrected by $\eff_s(\that)$
is probably a reasonable test of the microlensing hypothesis.
Thus we find good consistency with the hypothesis that our 42 events
are due to microlensing (although, strictly speaking, it would be preferable
to do this test with full blend efficiencies).

Finally, we should also compare the $\umin$ distribution
for the 13 clump giant events with the theoretical distribution
modified by the sampling efficiency because this is the set of events
which we use to draw our conclusions regarding the microlensing optical depth
toward the Bulge. For these 13 events (shown in Figure \uminKSfig(b))
we find a K--S statistic of .046 which yields
$P_{KS} = $0.31. Thus, this data subset is also consistent with the
microlensing prediction of a uniform $\umin$ distribution.

\chapter{Optical Depth Estimates}

The results of a gravitational microlensing experiment can be described
in terms of the optical depth, $\tau$, or the detected event rate, $\Gamma$,
along with the set of event timescales. 
The optical depth is a measure of the mass in microlensing objects
along the line of sight to the source stars, and it has the virtue that
it is independent of the mass of the lensing objects as long as those
masses generate events which fall within our region of sensitivity.
Thus, one can compare predictions and measurements of $\tau$ using models
of galactic mass distributions, without requiring details of the
mass functions and velocity distributions of the microlensing objects.
Then, for models which can match the observed optical depth, one
can compare more detailed mass functions and velocity distributions
with the set of observed event timescales, $\{ \that_i \}$.

Experimentally, one can define an estimated optical depth as the observed
microlensing rate times the efficiency weighted average event duration: 
$$ \tau_{\rm est} = {\pi \over 4\,E}
                     \sum_i {\hat t_i \over \eff(\hat t_i)}\ ,
\eqn\taum $$
where $E$ is the total exposure (in star-years), $\hat t_i$ is the
fit Einstein ring diameter crossing time for the $i$-th event,
and $\eff({\hat t_i})$ is the
detection efficiency as a function of $\that$.

One difficulty with using the optical depth rather than the event rate
to quantify the `amount' of microlensing is that since each event
contributes a different amount to the optical depth (eq. \taum),
the uncertainty in the optical depth does not follow Poisson statistics.
However, since the number of events still obeys Poisson statistics, it
is straight forward to evaluate confidence level limits using Monte
Carlo simulations in which the number of events for each simulated experiment
is selected according to
Poisson statistics. In order to calculate the optical depth for a given
simulation, each simulated event must also be assigned a timescale,
$\that_i$. The timescales are selected randomly from an assumed
distribution of event timescales. If the number
of detected events, $N$, is large, one can simply use the observed set of
event timescales, $\{ \that_i \}$. However, because both the mean and variance
are measured from the same dataset, this procedure will underestimate the
variance of the timescale distribution by a factor of $(N-1)/N$. Nevertheless,
we shall use this procedure for estimating the error bars in the present
paper because it less dependent on theoretical models. The reader should
be warned that the optical depth errors are being underestimated for
data subsets with small numbers of events.

The `raw' bulge optical depth estimated for our full sample using
sampling efficiencies $\eff_s(\that)$ is
$\tau_{\rm all,raw} = 1.46 \ten{-6}$. (The full sample includes 41 events:
the 43 which pass the cuts minus the 2 events which are probably not
microlensing.) However, this value for $\tau$ is almost
certainly an underestimate because of blending effects, and the
fact that some fraction of the source stars are foreground stars
in the foreground disk for which the optical depth is considerably lower.
As we have discussed
above, the use of sampling efficiencies is a rather poor approximation
for the fainter stars in the sample. Toward the LMC, we found that the
ratio of the full efficiencies to the sampling efficiencies was
$\eff_b(\that)/\eff_s(\that)\approx 0.66$--$0.8$ depending on $\that$.
Toward the bulge, this difference might be larger because
the average stellar density is higher, and
we are detecting more photons from the faintest stars we can identify
toward the bulge. (Our limiting magnitude is determined by crowding.)
This means that crowding-related systematic errors will be larger
relative to the random errors in the bulge.
We are undertaking a ``full'' efficiency analysis similar to that
for the LMC, but for the present we just apply a crude correction
factor of
\def\fblend{f_{\rm Blend}}
$\fblend^{-1}$ where we estimate
$\fblend \sim \langle \eff_B (\that) \rangle /
  \langle \eff_S (\that) \rangle \approx 0.75$.

The second important effect that will lower our raw $\tau$
value is that a significant fraction of the stars seen in these fields
are probably foreground disk stars for which the microlensing optical depth
would be lower.
In principle, we could also have background disk stars
which would have a higher optical depth, but our lines of sight toward
the Bulge are several hundred parsecs out of the Galactic plane on the
far side of the Bulge. Also, in the very crowded fields
we have observed, we expect that most of the disk `contamination' will be
from foreground stars. As we will see in the next section, there is some
independent evidence that the source stars for clump giant and
`non-clump giant' events come from different populations as the timescale
distributions of the two samples are significantly different from one
another.
The fraction of disk stars in our fields is rather
uncertain, but Minniti (1995) estimates from the DIRBE maps
of Weiland \etal\ (1994)
that the disk contributes $15\%$ of the integrated $2.2$ micron flux
from Baade's Window.
Since our magnitude limit is fainter than the base of the
giant branch, most of the optical flux should be resolved into stars in
our data.
Since the disk stars are on average
bluer and will suffer less extinction, the fraction of disk stars in our
optical passbands is probably larger than their fractional
contribution to the 2.2 micron flux; thus
\def\fdisk{f_{\rm Disk}}
we adopt a conservative estimate $\fdisk \approx 0.2$, and then correct our
estimated optical depth by  a factor of $(1 - \fdisk)^{-1}$ .

Then, our corrected estimate for the optical depth for bulge stars
using our full sample of 41 events is
 $\tau_{\rm all,cor} \approx 2.4 \pm 0.5 \ten{-6} \,
  (0.75 / \fblend ) \, (0.8 / (1 - \fdisk)) $,
where the latter two factors are somewhat uncertain but
should be fairly close to 1.

\section{Optical Depth for Clump Giants}

As we have seen above, there are significant uncertainties
in our estimated bulge optical depth from the full sample,
both due to blending
and the fact that some of the stars seen
towards the galactic bulge are foreground disk stars.
One way to avoid both these problems is to concentrate on a class of stars
that is both bright and `known' to be in the galactic bulge: the `clump
giant' stars. These are relatively low mass core helium burning giants--the
horizontal branch of a metal rich population, and this is the same population
used by the OGLE collaboration (Stanek \etal\ 1994; \pac\ \etal\ 1994)
to find evidence for
a Galactic bar in their data. The location of the clump giants is marked on
the color magnitude diagram in Figure \cmdfig. Note that due to the variable
extinction across our fields, the position of the giant clump is spread
out in the color magnitude diagram shown in Figure \cmdfig, but fortunately
the clump stars still occupy a distinct region of the color magnitude
diagram.

Table 2 shows our estimated optical depths and various confidence level
limits for a number of different cuts on the data. As mentioned above,
it is the `clump giant' values that are the most reliable measurements
of the optical depth toward stars in the bulge. Thus, our estimated optical
depth toward the clump giants
\footnote\dag{The $\tau$ for the clump giant sample differs from that in a
previous paper (Bennett \etal\ 1995) because the clump giant cut used
previously was too loose and included many stars which are not clump giants.}
averaged over all our fields is
$\tau_{\rm CG} = 3.9 {+ 1.8 \atop - 1.2} \times 10^{-6}$.
This is somewhat higher, but only at the $\sim 1\sigma$ level,
than the corrected optical depth for all stars estimated above:
 $\tau_{\rm all,cor} \approx 2.4 \pm 0.5 \ten{-6}
  (0.75 / \fblend) (0.8 / (1 - \fdisk) $.

There are several factors that could contribute to this difference. First,
it could be that $\fblend < 0.75$ or $\fdisk > 0.2$ since our estimates of
these values are very rough. Another important consideration is that the
number of stars observed across our fields is limited by crowding for the
full sample but not for the clump giant sample. This means that the number
density of stars in the full sample does not vary much across the sky
while the density of clump giants varies by a factor of 6 across our fields.
This is the reason that the average $\ell$ and $b$ values for the two
samples shown in Table 2 are different. Since the clump giants are more
concentrated toward the Galactic center, we should expect a larger optical
depth for this sample. We should also expect that the clump giants are
systematically more distant than the full sample of bulge stars. Faint stars
on the far side of the bulge are less likely to be detected because they
appear fainter than similar stars on the near side of the bulge, but
essentially all of the clump giants are bright enough to be in our sample, so
the nearer ones are less favored. This effect also serves to increase the
optical depth seen towards the clump giants.
In any case, the difference in optical depth seen toward these two samples
does not have a large statistical significance, and
the implied $95\%$ lower limits on the optical depth
are quite similar for the two samples. We will use the clump giant sample
for our main conclusions because the uncertainties are better understood for
this sample.

\section{Optical Depth as a Function of Latitude}

The third and fourth rows of Table 2 show $\tau_{\rm bulge}$ for two
different cuts on Galactic latitude:
$\vert b\vert < 3.5^\circ$ and $\vert b\vert > 3.5^\circ$. Although, the
majority of stars we have observed fall into the $\vert b\vert > 3.5^\circ$
category, the density of clump giants falls off quite rapidly away from the
galactic plane and the observed population of clump giants is about evenly
divided about $b = -3.5^\circ$. It was originally hoped that the variation
of $\tau_{\rm bulge}$ with galactic latitude might help to distinguish
different Galactic models which seek to explain the observed microlensing
optical depth toward the bulge. Certainly, the strong variation of $\tau$
with $b$ is what one would expect from lensing by the disk, but unfortunately,
the situation is not very different with models where the Galactic bar is
the dominant source of lenses (Zhao \etal\ 1994; Han \& Gould 1995a).
Thus, the apparent variation of $\tau_{\rm bulge}$ with $b$ does not
resolve the question as to whether the bulk of the lenses are in the
Galactic disk or bar.
We are continuing our observations and have
extended our coverage in both Galactic latitude and longitude in order
to obtain improved constraints on the latitude and longitude dependence
of $\tau$ and the microlensing event rate: in particular,
we are now monitoring a number of fields at $l \sim 10^o, b \sim -3^o$
where disk lensing is expected to dominate.
We will present results of these
investigations in a future publication.

\section{Comparison of Optical Depth with Model Predictions and Other Data}

The predictions of the microlensing optical depth toward the Galactic bulge
made prior to the first experimental estimates were in the range
$\tau \approx 1-1.5\times 10^{-6}$ (Griest \etal\ 1991; \pac\ 1991;
Kiraga \& \pac\ 1994) toward Baade's Window. This is about a factor of
3 below the best fit value $\tau_{CG} = 3.9\times 10^{-6}$ reported
here and below our 95\% confidence level lower limit of $1.9\times 10^{-6}$.
We note that these values are averaged over $\sim 10$ square degrees
centered at $\VEV{\ell} = 2.55^\circ$ and $\VEV{b} = -3.64^\circ$; this
is close to but not the same as Baade's Window at $\VEV{\ell} = 1.0^\circ$
and $\VEV{b} = -3.9^\circ$. In most models, the optical depth averaged over
our fields is very close to the optical depth at Baade's Window, so we
ignore this distinction for the rest of this section.

The results presented here agree quite well with our previous
result based on a subset of this data (Alcock \etal\  1995a)
and with the OGLE result of
$\tau_{\rm BW} \geq 3.3 \pm 1.2 \times 10^{-6}$ toward Baade's window
(Udalski \etal\ 1994a).
In comparing to this value, it is often not realized that this value
is a {\it lower limit} on the optical depth because they have ignored
blending effects when estimating their detection efficiencies. (The
OGLE paper makes this point quite clearly.) Since
they do not restrict themselves to a bright subset of stars, the blending
correction may be substantial. For example, if their
data and analysis were similar to our LMC data and analysis, then their
estimated optical depth towards Baade's window would increase by a factor
of $\sim 1.25$ when blending is taken into account. However, we should
emphasize that the differences between our data sets and
analysis methods might imply that the correction factor for OGLE would
be quite different from this.

In response to the first optical depth estimates toward the bulge, a number
of attempts have been made to produce models which can account for the
microlensing seen toward the bulge. We (Alcock \etal\ 1995a) and
Gould (1994) have suggested that the galactic disk might contain most
of the Galactic mass interior to the solar circle. For a double
exponential disk with any scale length and height,
the optical depth toward the center of our fields is
$\tau \lsim 1.6 \times 10^{-6} (v_d/200\ {\rm km/sec})^2$, where $v_d$ is the
circular velocity at the solar radius due to the disk. If we add the
contribution, $\tau = 5\times 10^{-7}$ due to a low mass axially symmetric
bulge to the disk value for $v_d = 200 {\rm km/sec}$, then the total,
$\tau = 2.1\times 10^{-6}$ is formally ruled out by our data at the 93\%
confidence level. A more extreme set of models was considered by
Gould who finds that $\tau \leq 3\times 10^{-6}$ at Baade's window for any
disk distribution with an exponential scale height and arbitrary
radial profile. Thus,
it is possible to have optical depths much closer to the measured value
with a massive disk that does not follow the exponential density relation
with radius.

Some have argued against the heavy disk models on the grounds that
the measured column density of the disk may be as small as
$50\ \msun {\rm pc}^{-2}$ (Kuijken and Gilmore 1989) although this estimate
is sensitive to the assumed model of the dark halo. Even if the
assumed halo model is correct, however, one must also take the
assumed exponential scale length dependence of the disk quite seriously
in order to make any connection between the local column density and the
global mass distribution of the disk. In fact, while the scaling of the
average disk mass density with radius seems to be well fit by an
exponential, the variation of the column density with galactic longitude
seems to be quite significant: a factor of 2--3 in cases that have
been studied in some detail (Rix \& Zaritsky 1995, Gnedin, Goodman \&
Frei 1995, Rix \& Rieke 1993). Thus, the apparently small local column density
is not a serious objection to the heavy disk model. Recent estimates
of the disk scale length which give values as small as $2.5\kpc$
(Fux \& Martinet 1994) also tend to indicate that a large disk mass
is compatible with the limits on the local disk column density.

Another type of model that has been proposed to explain the high optical
depths seen toward the bulge is a Galactic bar with the long axis
pointing close to the line of sight.
The existence of a bar has been suggested by
a number of authors (de Vaucouleurs, 1964; Blitz \& Spergel 1991 and
references therein), although often
it was predicted to be inclined by $\sim 30^\circ$. If the bar is
pointing near the line of sight, it can be considered to be an
`efficient' structure for generating microlensing because the mass in
such a bar is concentrated along the line of sight toward the source stars.
\pac\ \etal\ (1994) suggested that a bar with a small inclination angle
might provide the large observed optical depth,
and Zhao \etal\ (1995) have developed a detailed bar model. Toward
Baade's Window, they find
an optical depth of $\tau = 1.7\times 10^{-6}$ due to lensing by the
bar and $\tau = 5\times 10^{-7}$ due to lensing by a truncated disk.

We have seen in a previous section that this truncated disk seems to be
in conflict with our observations of microlensing of stars on the
upper main sequence, but Zhao, \etal\ also quote an optical depth of
$\tau = 6-9\times 10^{-7}$ for an untruncated disk. Comparing with
our optical depth results,
we find that the total $\tau$ of $2.2\times 10^{-6}$ for the
truncated disk model is formally ruled out at the 93\% confidence level while
if we take the maximum, untruncated disk model the total is
$\tau = 2.6\times 10^{-6}$ which can be formally excluded only at
85\% confidence.

A model with a fairly massive disk and a massive axially symmetric bulge
has been proposed by Evans (1995). He assumes that a rather large fraction
of the stars observed are actually in the disk, but because we are considering
only the clump giant sources, we will use his numbers for lensing of
bulge stars. Similarly, we will also ignore the contribution from lensing
by halo objects since it is unlikely that the Galaxy has a very massive
halo composed of substellar Machos (Alcock \etal\ 1995c, 1995d; Aubourg
  \etal 1995).
Toward Baade's window,
the Evans model gives an optical depth of $\tau = 1.9\times 10^{-6}$
with half the contribution from disk lensing and half from bulge
lensing. This value is formally excluded at the 95\% confidence level.

Finally, we should mention one caveat to the optical depth comparisons
mentioned in this section. Most of the calculations of the optical depth
due to lensing by the bulge or bar have assumed that we have a magnitude
limited sample of stars so that we can see a larger fraction of the
stars near the front of the bulge than toward the back. For the clump
giant sample considered here, a better assumption would be that
our source stars represent a complete set of stars that can be
seen throughout the bulge. The adoption of this assumption would increase
the predictions of some of the models mentioned above, but the main conclusion,
that the most of the recently proposed models can marginally fit our
optical depth measurement should remain unaffected by this correction.

However, if further data reveals that the true optical depth is in fact
close to or higher than our central value, this would prove rather
hard to account for  with the models proposed to date.

\chapter{Microlensing Event Timescales}

In addition to the microlensing optical depth, the distribution of
microlensing event timescales can also be used to constrain models.
However, unlike the optical depth, the timescale distribution depends on
the mass function and the spacial and velocity distributions
of the lensing population. The formula
for the Einstein Ring diameter crossing time is
$$ {\hat t} = {2\over v_\perp} \sqrt{4Gm L x(1-x)\over c^2 } \ ,
\eqn\thateqn $$
as can be seen from eq. \rEeqn.

Figure \thatfig\ is a histogram of the
$\that$ distribution of the 41 events which
pass the cuts used in the optical depth determination, while Figure \taufig\ is
the histogram for the same events with each event weighted by its
${\hat t}$ value which is proportional to its contribution to the optical
depth. Figures \thatfig\ and \taufig\ indicate while most of the events have
short timescales ($\lsim 40$ days), about a third of the contribution to the
optical depth comes from the long time scale events.
This might be taken as an indication
of microlensing by the disk since most bulge microlensing models do not tend to
predict many long timescale events (Kiraga \& \pac\ 1994; Evans 1995).
However, models with massive bars might also predict
a reasonable number of long timescale events if the pattern rotation speed of
the bar is large or if they contain a significant population of massive
objects such as neutron stars.

It is convenient to use the Kolmogorov-Smirnov test to compare the
observed distribution of ${\hat t}$ with model predictions, but the K-S
test can also be used to compare the ${\hat t}$ distributions for
different subsets of the data. Figure \thatcncfig\ shows the
cumulative distributions of event timescales for the clump giant events
compared to the other bulge events. It is clear that the clump giant events
have significantly longer time scales and the probability that this could have
occurred by chance is $P_{KS} = 0.007$. The reason for this difference is
not well understood at present, but there are several known factors that
contribute to this. First, the spatial distribution of the clump giants and
the `non-clump giants' is different. The clump giants are concentrated more
toward the Galactic Center at low Galactic latitudes. Thus, the clump
giants are somewhat more likely to be lensed by disk stars than the
`non-clump giants' which would be preferentially lensed by other Bulge
stars which would give them somewhat shorter timescales. However, the
difference between the center points of each of these distributions is
only about $0.3^\circ$ in latitude, so this seems unlikely to account for
all of the difference.

Another factor which may be somewhat more important is blending. The
clump giants are unlikely to be blended significantly, but the
`non-clump giants' tend to be much fainter and some of them
are probably seriously blended.
The fit $\that$ values for blended stars are often significantly
underestimated.
A final effect that might influence the $\that$ distribution is the
location of each set of stars along the line of sight. The clump giants
are generally located in the Bulge, but the `non-clump giants' include
a significant fraction of foreground disk stars. Thus, some of the
`non-clump giant' events may be due to disk-disk lensing. Unfortunately,
this last effect has the wrong sign to explain the difference between the
$\that$ distributions because the disk-disk lensing events should have
larger average $\that$ values.

Figure \Evansfig\ shows a comparison of the clump giant ${\hat t}$
distribution to the Evans (1995) model discussed above.
Evans assumes a Scalo (1986) mass function for the disk lenses and a
Richer-Fahlman (1992) mass function for the bulge.
The Richer-Fahlman mass function
rises very steeply for low masses and this gives rise to a great many
short timescale microlensing events. This model is clearly ruled out by the
K-S test which gives $P_{KS} = 0.0009$.
Of course, if we are free to choose
our mass functions, we could replace the assumed Richer-Fahlman mass function
for the bulge component of Evans model with something which fits our
observations, and so a test such as this cannot rule out bulge models unless
we are willing to assert that the implied mass function is unreasonable
on other grounds. The determination of the mass model by comparison to
the OGLE microlensing observations is the approach taken by Zhao \etal (1995),
so a similar test is not possible with their model.

\section{ Timescales for a Bar Model}

We have also estimated timescale distributions for the Han \& Gould
  (1995a) bulge model, for various mass functions.
 We use a barred bulge as in Han \& Gould, with a density profile
\def\pc {{\,\rm pc}}
 $$ \eqalign{
 \rho_B  & = 2.07 \, \exp(-w^2 / 2) \, \msun \pc^{-3} ,  \cr
     w^4 & = \left[ \left( {x' \over 1580 \pc} \right)^2
  + \left( { y' \over 620 \pc } \right)^2 \right]^2
   + \left( { z \over 430 \pc } \right)^4
 }
   \eqn\hgbareq
 $$
Here $x', y'$
 are measured along the axes of the bar in the galactic plane;
the $x'$ axis is  aligned $20^o$ from the GC-Sun line,
with the near side of the bar in the  positive-$l$ quadrant, and $z$ is
the usual height above the galactic plane.
 The total bar mass in this model is $20.65 \, \rho_0 \, abc$,
 where $a,b,c$ are the
 scale lengths above; i.e. $M = 1.8 \ten{10} \msun$ for our chosen
 values.

We use a double-exponential disk with a density profile
 $$ \rho_D = 0.08 \exp((R_0 - s)/3.5 \kpc)
    \exp(-\vert z \vert / 325 \pc )   \msun {\rm pc}^{-3} ,
  \eqn\diskeq
 $$
where $s, z$ are Galactocentric coordinates in cylindrical polar coordinates.

These models give optical depths towards Baade's Window of
  $1.3 \ten{-6}$ from the bar, and $0.6 \ten{-6}$ from the disk.

We assume that the velocity distribution function for the bar is
a Gaussian with a dispersion of $110 \kms$ in each transverse direction,
 and for the disk is a flat rotation curve of $220 \kms$ with
a Gaussian dispersion of $30 \kms$ in each direction.
We assume that sources reside in the bar, and are distributed in distance
 proportional to the local density (i.e. the luminosity function
 cancels the change in volume element, or $\beta = -1$ in the notation
 of Kiraga \& Paczynski (1994)).

Using a Monte-Carlo simulation with the above parameters,
we have evaluated the distribution of event timescales $\that$
for several illustrative mass functions:
delta-functions at $0.1$ and $1 \msun$,
 a Scalo (1986) main-sequence PDMF, and two power-law mass functions with
 $\phi(m) \propto m^{-\alpha}$:
one with $\alpha = -2.3$ for $0.1 < m < 1.4 \msun$, and one with
 a large number of brown dwarfs,
 with $\alpha = -2$ for $0.01 < m < 1.4 \msun$.
A broader class of models has been studied by Han \& Gould (1995c),
 with fairly similar conclusions to those below.

Results are shown in Figure~\thatmodfig\ for our sample of 41 bona fide
microlensing candidates which pass the cuts (excluding the 2 probable
variables). We include the detection efficiency $\eff(\that)$
in the model predictions, and normalize the model predictions to the
observed number of events.
Given the systematic uncertainties both due to crowding and the
fact that the bulge velocity structure is not  well known,
it is unwise to quote rigorous significance levels, but several
general conclusions can be drawn.

\item{(i)} Note that the range of model timescales is large even for
the delta-function mass functions. This means that it will be
 very hard to extract information about
fine details of the mass function. However, the mean mass of the
lenses can be fairly well constrained, and microlensing has
the major advantage that the estimate
is independent of the luminosities of the lenses.

\item{(ii)}
The observed range of timescales is roughly consistent
with most of the lenses being low-mass stars in the range $0.1 < m < 1 \msun$.
There is some evidence for an excess of short-timescale events over
the predictions from the Scalo (1986) PDMF; as noted by Han \& Gould
(1995c), a power-law with $\alpha = -2.3$ provides a better fit.
However, this could also possibly be explained by blending causing the
fitted event timescales to be systematically shortened.

\item{(iii)}
The mass function with a large mass fraction ($\approx 45\%$)
 of brown dwarfs predicts
many more short-timescale events ($\that < 10$ days) than observed;
thus, the disk and bulge mass are probably dominated by objects
 more massive than $0.1 \msun$, though a modest fraction of brown
 dwarfs would be allowed.

\item{(iv)}
There is marginal evidence for an excess of long-timescale
 events ($\that > 100$ days) relative to the Scalo and power-law
 PDMF's.
  This could be due to an additional population of massive lenses;
 ordinary stars with $m \simgt 1.5 \msun$ would be brighter than our
 typical source stars so are excluded,
 but a substantial population of neutron stars or black holes is possible.
 The statistical significance of this excess is not large,
 and it could also be explained by a population of slower-moving
 objects with a conventional mass function.
 Although more statistics are clearly desirable, these 4 long events
 are interesting because they contribute about a third of the estimated
 optical depth.

\chapter{Implications of Microlensed Main Sequence Stars}

Although Figure~\cmdfig\ indicates that the brighter main sequence stars
are less likely to be microlensed than giants of similar brightness,
the fact that {\it any} microlensing events are seen along the bright
main sequence is actually difficult to explain in some Galactic models.
Based on an analysis of the OGLE color magnitude diagrams,
\pac\ \etal\ (1994) have argued that the Milky Way disk has
a hole in the center starting 3-4 kpc from the solar radius. They also
argue that the stellar density through the disk stays nearly constant
along the line of sight to Baade's Window because the effect of the
rising exponential scale length tends to cancel the effect of the
falling exponential scale height. This would be a reasonable approximation
for our data as well because our average galactic latitude is close to that
of Baade's Window. With this assumption, it is straightforward to show
that the optical depth to disk lensing is 4 times larger for stars in the
Galactic bulge than for those in the disk.

Of course, bulge stars may be
lensed by other bulge stars, as well. Since a truncated disk has a small
microlensing optical depth, a Galactic bar pointed along the line of sight
is probably required to explain the microlensing results in these models.
In the bar model of Zhao, Spergel \& Rich (1994),
which assumes a truncated disk,
the disk only accounts for 20\% of the total optical depth. This implies that
the predicted optical depth toward the disk main sequence stars should be only
5\% of the optical depth toward the giants because the disk stars can only
be lensed by other disk stars and because the average distance to the
disk stars is about 1/4 of the distance to the giants in the Bulge.
Now for disk-disk lensing,
the source, lens, and observer are all rotating at nearly the same
velocity, so the value of $v_{\perp}$ should be very small, and the
$\that$ values should be larger than for lensing of bulge sources.
This would imply the
event rate for main sequence stars should be {\it smaller} than
5\% of the rate for the giants. As we discuss below, we find 13 events
in the 1.3 million stars which fall into our clump giant category. If
we select all the stars which fall to the left of the clump giant
region which are brighter than $\Vmacho = 17.5$, then we have
500,000 stars in this bright main sequence region. At 5\% of the observed
rate toward the giants, we would expect to find 0.25 events in this
category. Figure \cmdfig\ shows 3 events in this category, but one of these
events is 111-B which fails our final cuts because its $\tmax$  is
too late. The Poisson probability of finding 2 or more events when 0.25
are expected is only 2.6\%, so the model of Zhao, \etal, combined with the
OGLE truncated disk model would appear to be formally excluded.

More specifically, this analysis indicates that at least one of the
following should be true:

\item{1)} The bright `main sequence' seen towards the Bulge contains a
significant number of stars further away than 4 kpc. These could be
true main sequence stars or perhaps blue horizontal branch stars.

\item{2)} Disk lenses contribute significantly more than 20\% of the
microlensing optical depth toward the Bulge.

\chapter{Implications for the Dark Matter Halo}

The large microlensing optical depth seen toward the Galactic bulge was
not predicted by ``standard" Galactic models (Griest \etal\ 1991;
Paczynski 1991).
The models that have been
proposed to explain these microlensing results tend to require that the
mass of Galactic disk plus the bulge (or bar) is higher than has previously
been supposed (Kiraga and Paczynski 1994; Evans, 1995;
Zhao, Spergel and Rich 1995;
Han and Gould 1995a,b,c; Gates, Gyuk, and Turner 1995a,b,c).
If the mass of the disk plus bulge in the inner Galaxy is larger than
previously thought, then for a given acceleration at the Solar circle,
one would expect that the mass of the halo (the only remaining
component) would be smaller
in the inner regions of the Galaxy than previously thought.
This in turn would lead one to infer a higher Macho {\it fraction} in
the halo, for a given estimate of the microlensing rate toward
the LMC.
Thus, while the
microlensing surveys toward the LMC can accurately constrain the mass of
Machos along the line of sight toward the LMC (Alcock \etal\ 1995c, 1995d;
Aubourg \etal\ 1995) they cannot determine the
Macho {\it fraction} of the halo very accurately because the
mass distribution of the Galactic halo is not well known.

In a previous
paper, we considered our LMC results in some detail and gave examples of
halo models which are consistent with a heavy disk (Alcock \etal\ 1995d).
We showed that the most
likely Macho fraction of the halo can increase by as much as
$\sim 40$\%
when a `standard' halo model is replaced by a heavy disk galactic model
with a flat rotation curve. It may even be possible to
construct models which are consistent with a 100\% Macho halo and with
galactic mass estimates based on satellite galaxies
(e.g. Zaritsky \etal\ 1989) if the disk is heavy.  Thus, it will
be important to understand the source of the large microlensing optical
depth seen toward the Galactic bulge before the microlensing optical
depth toward the LMC can be accurately translated into a measure of
the Macho fraction of the Galactic halo.

Gates, Gyuk, \& Turner (1995a,b,c) have also considered
the implications of the bulge lensing results on the MACHO fraction of the
halo. They explored a large parameter space of models which comprised
various combinations of halo, disk, and bulge. Gates \etal\
differ from us by concluding that a {\it higher} microlensing
optical depth toward the Galactic bulge implies a {\it smaller} MACHO
halo fraction. This conclusion arises as follows: the {\it disk} model
in the Gates \etal\ simulations has a  double
exponential density profile, with a limit on the local column density.
The local column density limit means that any model which has a
massive disk will have a small radial scale length. When Gates \etal\
add a massive bar to a massive disk with a short scale length, they
find that this puts too much mass near the Galactic center, which in
turn produces a rotation curve inconsistent with observation in the
inner Galaxy. Thus, they find that a massive
bar {\it and} a massive disk are incompatible.

The conclusion of Gates \etal\ depends upon the
assumption of a exponential disk. The actual disk
density may differ substantially from the assumed exponential
radial dependence. The implications of the
microlensing results toward the bulge for the MACHO fraction of the halo
remain unclear. More information from microlensing, including
latitude and longitude dependence of the microlensing
optical depth, will be needed to resolve this issue.

\chapter{Future Developments}

As mentioned above, we are continuing our microlensing survey of the
Magellanic Clouds and the Galactic Bulge with significantly expanded
galactic latitude and longitude coverage. This will aid us in determining
the relative contribution to the microlensing optical depth from lenses
in the Disk and Bulge or Bar.

Near the end of the the 1994 Bulge observing season, we developed the
capability to detect events in progress (Alcock \etal\ 1994, Bennett \etal\
1995, Pratt \etal\ 1995).
This capability, which has also been demonstrated by the OGLE collaboration
(Udalski \etal\ 1994c) allows detailed and comprehensive studies
of the microlensing events in progress. For example, our
first real-time event led to the first spectral observations of
a microlensing event in progress (Benetti \etal\ 1995); the spectrum was
unchanged during the event,
a strong confirmation of
the microlensing hypothesis. We have discovered
about 40 microlensing events in progress in the 1995 observing season
(see {\tt http://darkstar.astro.washington.edu/} for details).
To take advantage of our real time detection capability, we now
operate the CTIO 0.9m telescope for 13\% of every night to
obtain more accurate photometry of ongoing events. This gives us a higher
sensitivity to detect `second order' microlensing effects (such as
parallaxes and caustic crossings) that will yield additional information
about the masses and distances of the lenses. It also gives us an
opportunity to confirm microlensing events which have poor light
curve coverage in the Stromlo data.

The high microlensing rate seen toward the Galactic Bulge coupled with the
ability to detect microlensing events in real time has caught the
attention of the planetary sciences community (Tytler \etal\ 1995), for
microlensing appears to be a very promising new approach for
detecting planets around distant stars
(Mao \& \pac\ 1991, Gould \& Loeb 1992).
Briefly, if every lensing star has a Jupiter-like planet,
then $\sim 15\%$ of microlensing events should show a deviation
$\delta A / A > 5\%$
from the single-lens microlensing light-curve.
Most such deviations would be quite short-lived ($\sim 1$ day), so
we could not reliably detect them in the present data;
but there appears
to be a fairly high probability that a serious microlensing follow-up
effort will soon commence to search for the binary microlensing signature
of a planet orbiting a lensing star.

\chapter{Summary and Conclusions}

We have presented the results of the first year MACHO Project
microlensing survey towards the Galactic Bulge. In this dataset consisting of
12.6 million stars observed for 190 days, we have discovered 45
candidate microlensing events. These include many
events with high amplifications and excellent signal-to-noise,
and two events showing exotic second-order effects:
one is the first observation of parallax in a microlensing event,
 and the other is a spectacular example of a binary lens
(first seen by OGLE).
The distribution of peak amplifications is consistent with
the microlensing prediction, and the events are distributed all across the
color magnitude diagram. The only reasonable interpretation of these data is
that the large majority of these events are, in fact, microlensing.

An unexpected feature of the color magnitude distribution of events is
the discovery of a couple of microlensed stars on the upper main sequence.
This is quite surprising because these stars had been thought
to be in the foreground of the vast majority of lenses. The existence of
these events if difficult to reconcile with Galactic models
(such as that of Zhao \etal\ (1995)) where $\lsim 25$\% of the microlensing
optical depth is due to the disk unless a substantial fraction of the bright
main sequence stars are more distant than 4 kpc.

For our determination of the microlensing optical depth toward the bulge,
we have focused on a subset of 13 events in which the lensed stars are
`clump giants\rlap.' These stars have the advantage that they are generally
located in the Bulge with little contamination from foreground disk stars
unlike the fainter stars in our sample. Another advantage of the clump
giant sample is that they are bright enough so that the blending of
stellar images does not have a significant effect on the microlensing
detection efficiencies. With this sample of 13 events, we find a microlensing
optical depth of
$\tau_{\rm CG} = 3.9 {+1.8 \atop -1.2} \times 10^{-6}$. This is
quite consistent with the OGLE result but marginally higher than the
most recent theoretical predictions.
Using our full sample of events,
 we find $\tau_{\rm all} \approx 2.43 {+0.54 \atop -0.45} \ten{-6}
  \, (0.75 / f_{Blend}) \, (0.8 / (1 - f_{Disk})) $, where $f_{Blend}$
 gives the degradation of our detection efficiency due to stellar blending,
 and $f_{Disk}$ is the fraction of our source stars in the foreground disk.

If the true optical depth is close to
our central value, then the inner Galaxy probably contains more mass than
in most Galactic models, and the dark halo may have a larger core radius
than is usually supposed.

\singlespace
\centerline{\bf Acknowledgements}
We would like to thank Wyn Evans for kindly providing the detailed timescale
distributions for his galactic bulge and disk model.
We are grateful for the support given our project by the technical
staff at the Mt. Stromlo Observatory.  Work performed at LLNL is
supported by the DOE under contract W-7405-ENG.  Work performed by the
Center for Particle Astrophysics personnel is supported by the NSF
through AST 9120005.  The work at MSSSO is supported by the Australian
Department of Industry, Science and Technology.
K.G. acknowledges support from DoE OJI, Alfred P. Sloan, and Cotrell Scholar
awards.
C.S. acknowledges the generous support of the Packard and Sloan Foundations.
\bigskip

\endpage
\centerline{\bf References}
\def\apj{{\it ApJ}}
\def\apjl{{\it ApJL}}
\def\annrev{{\it Ann. Rev. Astron. Astrophys.}}
\def\mn{{\it MNRAS}}
\def\aap{{\it AAp}}
\def\ref{\par\hangindent=1cm\hangafter=1\noindent}
\parskip 0pt

\ref Alcock, C., Akerlof, C.W., Allsman, R.A., Axelrod, T.S., Bennett, D.P.,
 Chan, S., Cook, K.H., Freeman, K.C., Griest, K., Marshall, S.L., Park, H.-S.,
 Perlmutter, S., Peterson, B.A., Pratt, M.R., Quinn, P.J., Rodgers, A.W.,
 Stubbs, C.W., \& Sutherland, W., 1993, {\it Nature}, {\bf 365}, 621.

\ref Alcock, C. \etal, 1994. {\it IAU Circulars} 6068, 6095.

\ref Alcock, C. \etal,
       1995a, \apj, {\bf 445}, 133 

\ref Alcock, C. \etal,
          1995b, \apj, {\bf 449}, 28  

\ref Alcock, C. \etal, 1995c, {\it Phys. Rev. Lett.} {\bf 74}, 2867

\ref Alcock, C. \etal, 1995d, \apj, in press. 	 

\ref Alcock, C. \etal, 1995e, \apjl, {\bf 454}, L125.  

\ref Alcock, C. \etal, 1995f, in preparation 


\ref Aubourg, E., Bareyre, P., Brehin, S., Gros, M., Lachieze-Rey, M.,
  Laurent, B., Lesquoy, E., Magneville, C., Milsztajn, A., Moscosco, L.,
  Queinnec, F., Rich, J., Spiro, M., Vigroux, L., Zylberajch, S., Ansari, R.,
  Cavalier, F., Moniez, M., Beaulieu, J.-P., Ferlet, R., Grison, Ph.,
  Vidal-Madjar, A., Guibert, J., Moreau, O., Tajahmady, F., Maurice, E.,
  Prevot, L., \& Gry, C., 1993, {\it Nature}, {\bf 365}, 623.

\ref Aubourg, E. \etal, 1995. \aap, {\bf 301}, 1  

\ref Bahcall, J.N., 1986. \annrev, {\bf 24}, 577.

\ref Bahcall, J.N, Flynn, C., Gould, A., \& Kirhakos, S. 1994, \apjl,
	{\bf 435}, L51 

%
\ref Bennett, D.P.  \etal, 1995, AIP Conference Proceedings 336:
     {\it Dark Matter}, S.~S.~Holt, C.~L.~Bennett, eds., p. 77.

\ref Bennett, D.P. \etal, 1996, in preparation   

\ref Benetti, S., Pasquini, L. \& West, R., 1995, \aap, {\bf 294}, L37.

\ref Blitz, L \& Spergel, D.N. 1991, \apj, {\bf 379}, 631

%

%
%
%

\ref de Vaucoulers, G., 1964, IAU Symposium 20: The Galaxy and the Magellanic
   Clouds, eds., F.J.Kerr \& A.W.Rodgers (Sydney: Australian Academy of
   Science), p. 195.

\ref Evans, N.W., 1995, \apjl, 437, L31

%
%

\ref Fux, R., \& Martinet, L., 1994, \aap, {\bf 287}, L21

\ref Gates, E.I., Gyuk, G. \& Turner, M.S., 1995a, \apjl, {\bf 449}, L123

\ref Gates, E.I., Gyuk, G. \& Turner, M.S., 1995b, {\it Phys. Rev. Lett.}
     {\bf 74}, 3724.

\ref Gates, E.I., Gyuk, G. \& Turner, M.S., 1995c, preprint.


\ref Gnedin, O.Y, Goodman, J., \& Frei, Z. 1995, astro-ph/9501112

%
\ref Gould, A., 1994, astro-ph/9408060  

\ref Gould, A., \& Loeb, A., 1992, \apj, {\bf 396}, 104 

\ref Gould, A., Miralda-Escude, J., \& Bahcall, J.N., 1994.
	\apjl, {\bf 423}, L105.  


%
\ref Griest, K., \etal\ 1991, \apjl, {\bf 372}, L79. 


\ref Han, C. \& Gould, A., 1995a, \apj, {\bf 447}, 53 

\ref Han, C. \& Gould, A., 1995b, \apj, {\bf 449}, 521 

\ref Han, C. \& Gould, A., 1995c, astro-ph/9504078 

%
\ref Kiraga, M \& \pac, B. 1994, \apjl, {\bf 430}, L101

\ref Kuijken, K. \& Gilmore, G., 1989. \mn, {\bf 239}, 605. 

%

\ref Mao, S. \& \pac, B., 1991. \apjl, {\bf 374}, L37

\ref Marshall, S. \etal, 1994. In {\it Astronomy from Wide-Field Imaging},
   Procs. IAU Symp. 161, eds. H. MacGillivray \etal, Kluwer.

\ref Minniti, D., 1995. ESO Preprint 1103.

\ref \pac, B., 1986, \apj, {\bf 304}, 1.

\ref \pac, B., 1991, \apjl, {\bf 371}, L63

\ref \pac, B., Stanek, K., Udalski, A., Szymanski, M., Kaluzny, J.,
      Kubiak, M., Mateo, M. \& Krzeminski, W.,
     1994, \apjl, {\bf 435}, L113	

\ref Petrou, M., 1981. {\it Ph.D. thesis}, University of Cambridge.

\ref Pratt, M. \etal, 1995. In {\it Astrophysical Applications of
   Gravitational Lensing}, Procs. IAU Symp. 173,
     eds. C. Kochanek \& J. Hewitt, Kluwer.

\ref Refsdal, S., 1964. \mn, {\bf 128}, 295

\ref Richer, H.B. \& Fahlman G.G., 1992. {\it Nature}, {\bf 358}, 383

\ref Rix, H.-W. \& Rieke, M.J., 1993, \apj, {\bf 418}, 123

\ref Rix, H.-W. \& Zaritsky, D., 1995, \apj, {\bf 447}, 82

%
%
\ref Scalo, J.M., 1986, {\it Fund. Cosmic Phys.}, {\bf 11}, 1

\ref Stanek, K.Z., Mateo, M., Udalski, A., Szymanski, M.,
   Kaluzny, J. \& Kubiak, M.,
     1994, \apjl, {\bf 429}, L73  

\ref Stubbs, C., Marshall, S.L., Cook, K.H., Hills, R., Noonan, J.,
  Akerlof, C.W., Axelrod, T.S., Bennett, D.P., Dagley, K., Freeman, K.C.,
  Griest, K., Park, H.-S., Perlmutter, S., Peterson, B.A., Quinn, P.J.,
  Rodgers, A.W., Sosin, C., \& Sutherland, W.,
    1993, {\it SPIE Proceedings}, {\bf 1900}, 192

\ref Schechter, P.L., Mateo, M., \& Saha, A., 1993,
    {\it PASP}, {\bf 105}, 1342


\ref Tytler, D. \etal, 1995, in preparation.

\ref Udalski, A., Szymanski, M., Kaluzny, J., Kubiak, M., Krzeminski, W.,
   Mateo, M., Preston, G.W., \& \pac, B.,
   1993, {\it Acta Astronomica}, {\bf 43}, 289   

\ref Udalski, A., Szymanski, M., Stanek, K., Kaluzny, J.,
    Kubiak, M., Mateo, M., Krzeminski, W., \pac, B. \& Venkat, R.,
      1994a, {\it Acta Astronomica}, {\bf 44}, 165   

\ref Udalski, A., Szymanski, M., Mao, S., Di Stefano, R.,
  Kaluzny, J., Kubiak, M., Mateo, M. \& Krzeminski, W.,
    1994b, \apjl, {\bf 435}, L113   

\ref Udalski, A., Szymanski, M., Kaluzny, J., Kubiak, M.,
     Mateo, M., Krzeminski, W. \& \pac, B.,
    1994c, {\it Acta Astronomica}, {\bf 44}, 227   


\ref Weiland, J.L. \etal, 1994. \apjl, {\bf 425}, L81.  

%

\ref Zaritsky, D., Olszewski, E.W., Schommer, R.A., Peterson, R.C.
  \& Aaronson, M., 1989. \apj, {\bf 345}, 759. 


\ref Zhao, H., Spergel, D.N., \& Rich, M. 1995, \apjl, 440, L13

\vfill
\eject

\figout
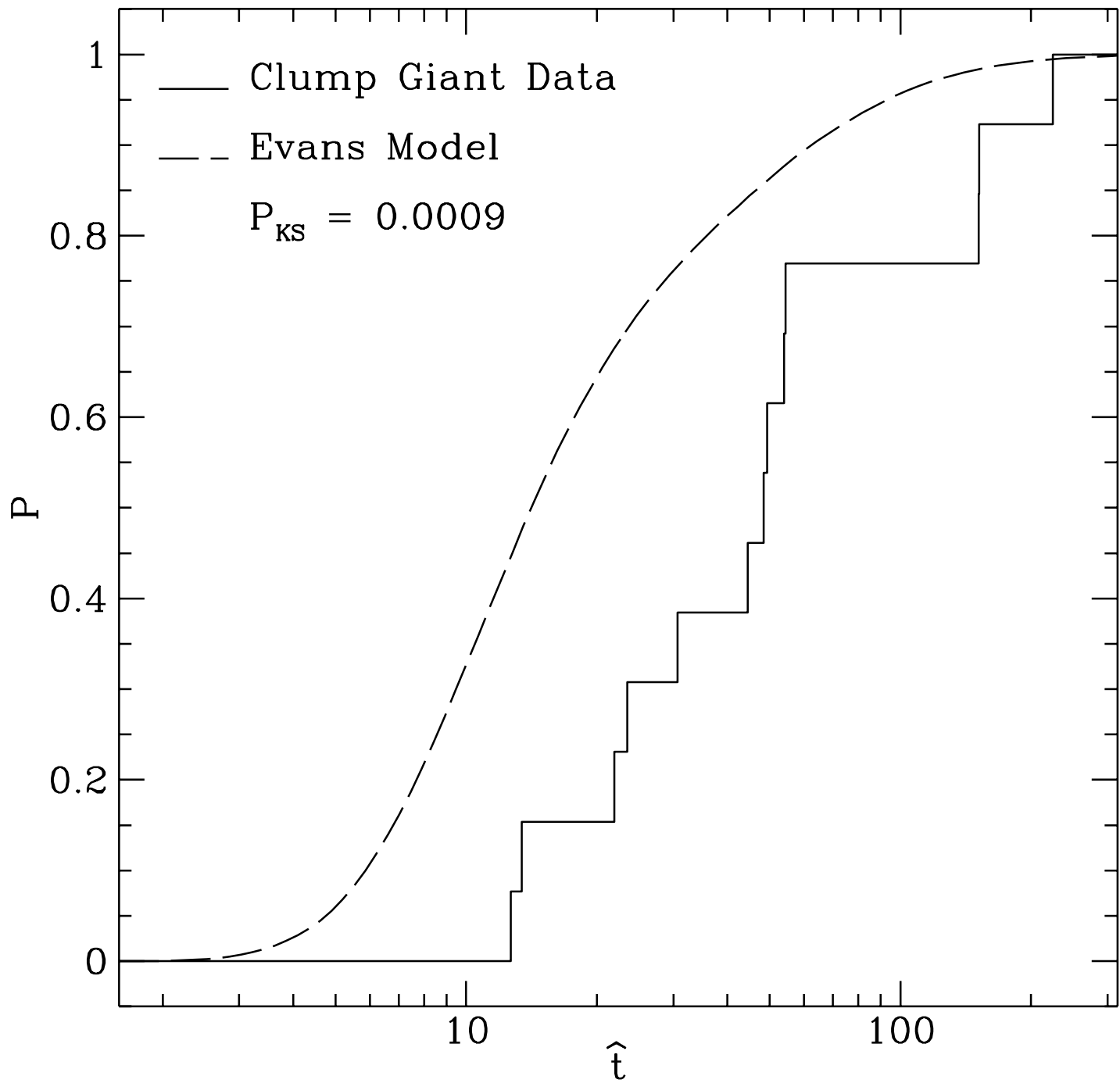
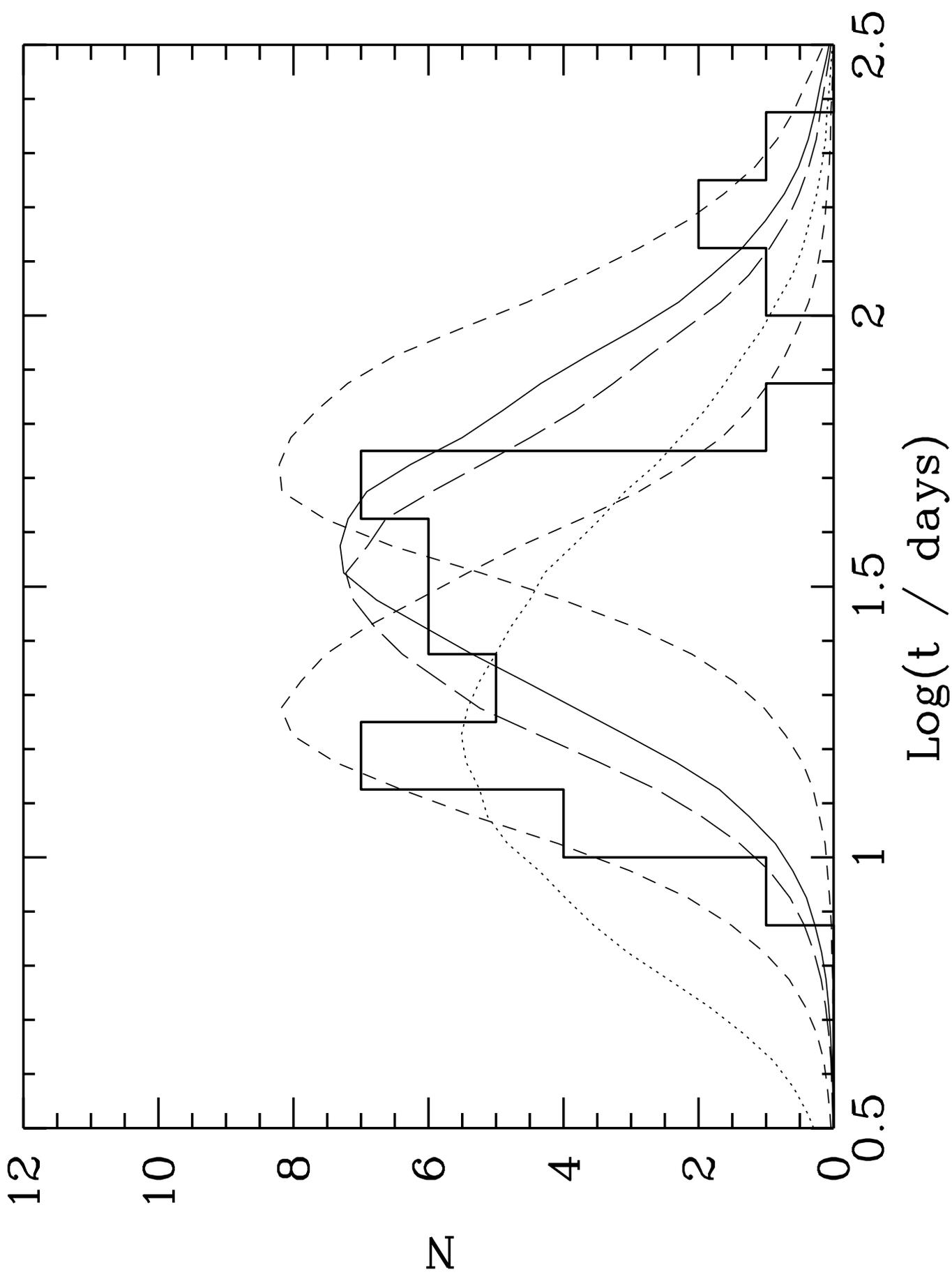
\end